\documentclass[doublecol]{epl2} 
\usepackage{graphicx}
\usepackage{float}
\usepackage{braket}
\usepackage{physics}
\usepackage{slashed}

\title{Cross-sections and experimental signatures for \\ detection of a well-defined dark matter WIMP}
\shorttitle{detection of a well-defined dark matter WIMP} 

\author{Bailey Tallman\inst{1} \and Jehu Martinez\inst{1} \and Rohan Shankar\inst{1} \and Kane Rylander\inst{1} \and Roland E. Allen\inst{1}}
\shortauthor{Bailey Tallman \etal}

\institute{                    
  \inst{1} Physics and Astronomy Department, Texas A\&M University, College Station, Texas 77843, USA
}

\abstract{
We report the following calculations for a recently proposed bosonic dark matter WIMP with well-defined interactions: (1)~the mass as determined by fitting to the relic abundance; (2)~the current annihilation cross-section for indirect detection; (3)~cross-sections for pair production accompanied by jets in proton colliders with center-of-mass energies ranging from 13 to 100 TeV; (4)~for the high-luminosity LHC, and planned 100 TeV proton collider, detailed plots of experimentally accessible quantities before and after optimal cuts; (5)~cross-sections, and plots of experimentally accessible quantities, for production in e$^+$e$^-$ or muon colliders with center-of-mass energies up to 10 TeV; (6)~cross-section per nucleon for direct detection. The conclusions are given in the text, including the principal prediction that (with optimal cuts) this particle should be detectable at the high-luminosity LHC, perhaps after only two years with an integrated luminosity of 500 fb$^{-1}$.}

\begin{document}

\maketitle

\section{\label{sec:sec1}Introduction}

In a previous letter~\cite{DM2021}, and subsequent papers~\cite{DM2021a,DM2022,DM2023}, we proposed a bosonic dark matter WIMP with precisely-defined interactions which are weak and second-order, implying low cross-sections that are consistent with experiment. Here we report much more extensive calculations relevant to its experimental detection within the near and more distant future~\cite{pdg-DM,Snowmass1,Snowmass2,Snowmass3,P5,review}. 

All these calculations were performed with the interactions in Eq.~(47) of \cite{DM2021}, using MicrOMEGAs~\cite{micrOMEGAs}, MadGraph~\cite{MadGraph}, and MadAnalysis~\cite{MadAnalysis}.

\section{\label{sec:sec2}Annihilation cross-section and mass from relic density}
\begin{figure}
\begin{center}
\resizebox{\columnwidth}{!}
{\includegraphics{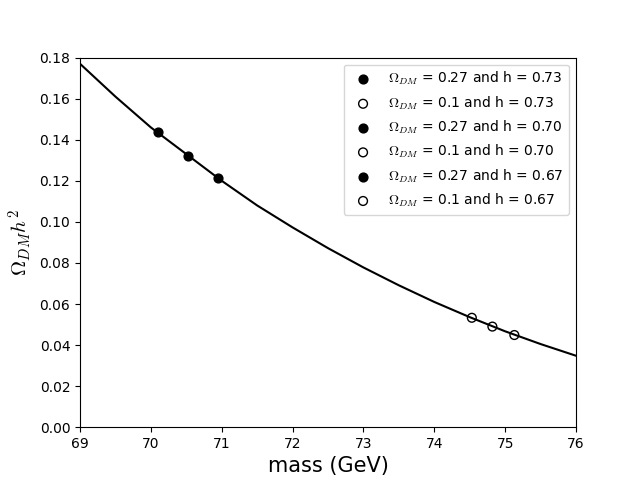}}
\end{center}
\caption{$\Omega_{DM} h^2$  
versus mass of the present dark matter WIMP. For the three points on the left, with a dark matter fraction of 27\% and three possible values of the Hubble parameter $h$~\cite{Hubble-tension}, the higgson is assumed to comprise most of the dark matter. The mass is then near 70 GeV in each case. For the points on the right the higgson is taken to be  subdominant in a multicomponent scenario, contributing only a 10\% fraction of the observed mass-energy content of the universe. The mass is then still not far from 70 GeV, and this is true even for substantially smaller fractions. The reason is that this particle would be extremely subdominant if the mass were above 80 GeV, permitting rapid annihilation into W bosons.}
\label{relic}
\end{figure}

Fig.~\ref{relic} shows our results (obtained with MicrOMEGAs) for the variation of the standard parameter $\Omega_{DM} h^2$ with the mass of the present dark matter particle,  which is designated $h^0$ and called a higgson as in our previous papers, because of its close relationship to the observed Higgs boson $H^0$. 

Here  $\Omega_{DM} =\rho_{DM} / \rho_c$, where $\rho_{DM}$ is the current density of dark matter and $\rho_c$ is the critical density. Also, $h=H_0$/(100 km s$^{-1}$Mpc$^{-1}$), where $H_0$ is the current value of the Hubble parameter.

If the present particle accounts for all or the great majority of the dark matter, it can be seen that its mass is close to 70 GeV for all values of $h$ within the range of currently accepted values~\cite{Hubble-tension}. 

Even in a multicomponent scenario where it is a substantial component,  
it still must have a mass below the 80 GeV mass of the W boson, since otherwise it would undergo rapid annihilation in the early universe. 
This means that the results below for collider detection still hold approximately for all scenarios in which the present particle constitutes a significant fraction of the dark matter.

\begin{figure}
\begin{center}
\resizebox{\columnwidth}{!}
{\includegraphics{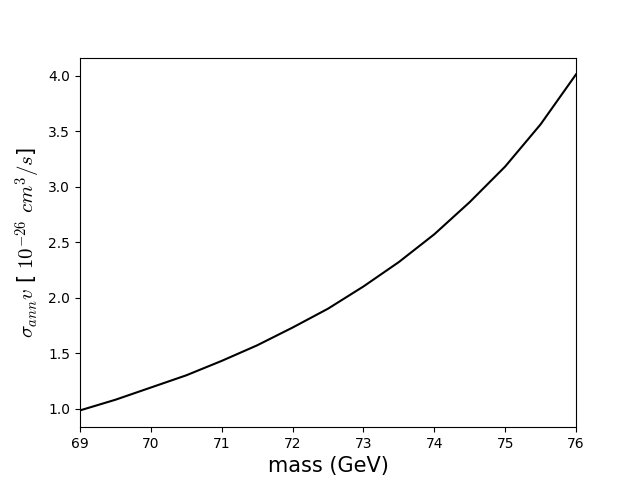}}
\end{center}
\caption{$\langle \sigma_{ann} v \rangle $ 
versus mass for the present dark matter WIMP. For the mass of 70 GeV established by the calculations shown in Fig.~\ref{relic}, $\langle \sigma_{ann} v \rangle  = 1.19 \times 10^{-26}$ cm$^3$/s. Larger masses would imply a smaller relic abundance, still consistent with the observational limits.}
\label{ann}
\end{figure}

The cross-section for annihilation in the present universe is shown as a function of mass in Fig.~\ref{ann}. For a mass of 70 GeV, our calculations yield a cross-section given by $\langle \sigma_{ann} v \rangle = 1.19 \times 10^{-26} $ cm$^3$/s. This value is consistent with the current limits from observations of dwarf galaxies~\cite{review,MAGIC,Calore-2020,VERITAS} (for a particle mass of 70 GeV and annihilation into e.g.  W$^+$W$^-$ pairs).
Our calculated mass and $\langle \sigma_{ann} v \rangle $ are also both consistent with analyses of the Galactic center gamma-ray excess observed by Fermi-LAT~\cite{Goodenough,Fermi,Hooper-gamma,Fermi-GCE,Leane-1,Leane} and the antiproton excess observed by AMS-02~\cite{Cuoco,Cui,Cuoco2,AMS-2,AMS-1}, as well as the multitude of observations by other experiments~\cite{IceCube,Planck,review}, although only the gamma-ray excess is currently regarded as potentially strong evidence of dark matter annihilation.

At one time it may have appeared that a positron excess from AMS-02 and other experiments was evidence for a dominant dark matter particle at an energy of $\sim 800$ GeV or above. However, this interpretation has been ruled out by Planck~\cite{Planck}, and the excess has been attributed to pulsars~\cite{Hooper-positron}. 

Among many other astrophysical implications, the present dark matter candidate is consistent with the formation of ``dark stars'' in the early universe~\cite{Freese}.

\section{\label{sec:sec3}Production cross-sections and experimental signatures in proton colliders}

\begin{figure}
\begin{center}
\resizebox{\columnwidth}{!}
{\includegraphics{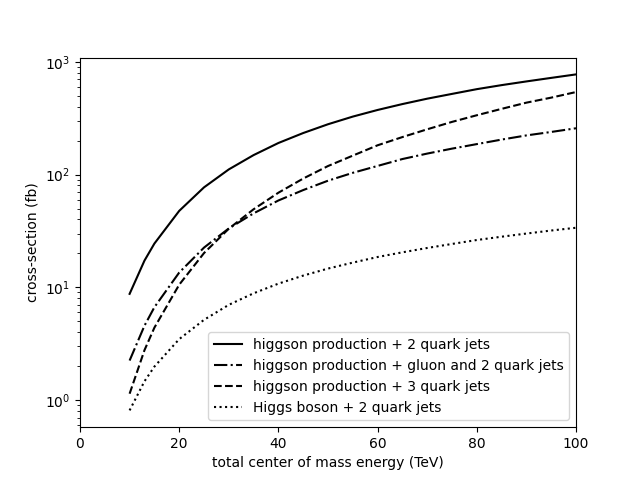}}
\end{center}
\caption{Cross-sections for production of the present particle -- called a higgson in our previous papers -- in proton colliders with center-of-mass energies $\sqrt{s}$ up to 100 TeV. For the 14 TeV high-luminosity LHC the cross-section is 20 fb. The similarly calculated cross-section for production of a Higgs boson pair through vector boson fusion is also shown.}
\label{cross}
\end{figure}

Using MadGraph and MadAnalysis, we have calculated cross-sections for production of the present dark matter particle $h^0$ in proton colliders with center-of-mass energies from 13-14 TeV (for the current and high-luminosity LHC) to 100 TeV (for the ultimate performance of the proposed future circular collider~\cite{FCC}).
The results for various signals are shown in Fig.~\ref{cross}. 

Fig.~\ref{cross} also shows our results for vector-boson-fusion (VBF) production of Higgs boson pairs, which are generally consistent with those of previous calculations~\cite{VBF-2013,VBF-2014,VBF2014,VBF2015,handbook,VBF-N3LO,report-2019,review-higgs,100TeV,VBF2017,VBF2018,pdg} (where it was found that the cross-section was sufficiently large for VBF to be a promising route to observing and measuring the $HHH$ and $HHVV$ interactions, with $H$ and $V$ respectively representing the Higgs and vector boson fields). 

\begin{figure}
\begin{center}
\resizebox{\columnwidth}{!}
{\includegraphics{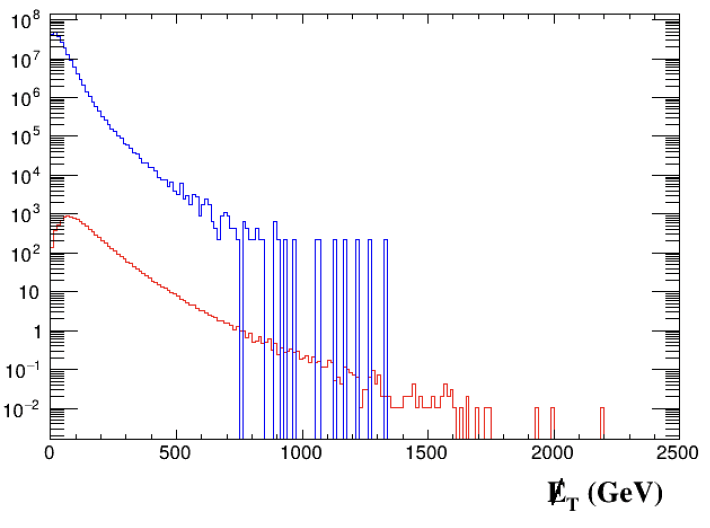}}
\end{center}
\caption{Anticipated number of events as a function of the ``missing transverse energy'' $\slashed{E}_T$ (as defined in \cite{MadGraph} and\cite{MadAnalysis}), for an integrated LHC luminosity of 500 fb$^{-1}$ and center-of-mass energy $\sqrt{s} =$ 14 TeV, prior to any cuts on the observables. (For all the histograms of this paper, 200 bins were used.) Here and in Figs.~\ref{5}-\ref{7}, the higher histograms are those for background. The calculations were done with $10^6$ events for the specified signal -- a pair of higgsons accompanied by 2 quark jets -- and $10^6$ events for the standard model background -- a pair of neutrinos accompanied by 2 quark jets.  The fluctuations in histogram height, here and in the plots below, are much larger for the background because the number of events was taken to be the same for signal and background in the calculations, but the total calculated cross-section is 20.6 fb for the signal and 436 pb for the background, so that the total anticipated number of events is about 21000 times larger for background, making the height for a single calculated event 21000 times larger.}
\label{4}
\end{figure}

\begin{figure}
\begin{center}
\resizebox{\columnwidth}{!}
{\includegraphics{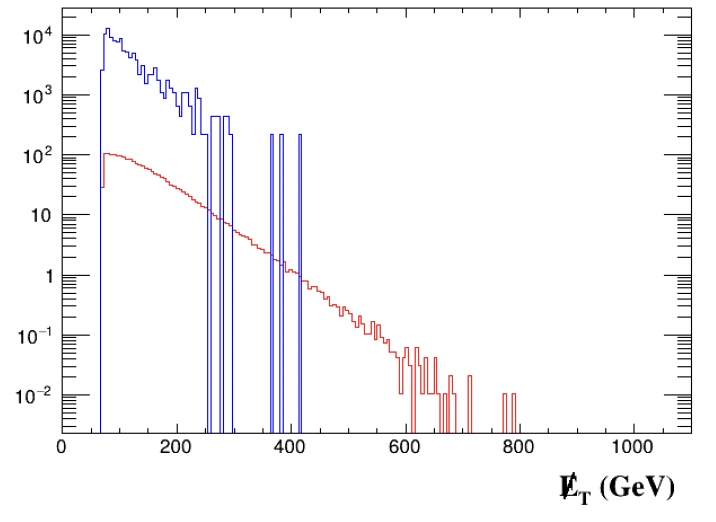}}
\end{center}
\caption{Anticipated number of events for the same sets of collisions as in Fig.~\ref{4}, but retaining only those events which satisfy the 9 cuts specified in Table 1. Experimentally, the ``signal'' events are seen as an excess over the predicted background, in the present case for 2 quark jets with missing transverse energy, after the cuts. One anticipates cuts much more sophisticated and effective than these initial attempts, of course.}
\label{5}
\end{figure}

\begin{figure}
\begin{center}
\resizebox{\columnwidth}{!}
{\includegraphics{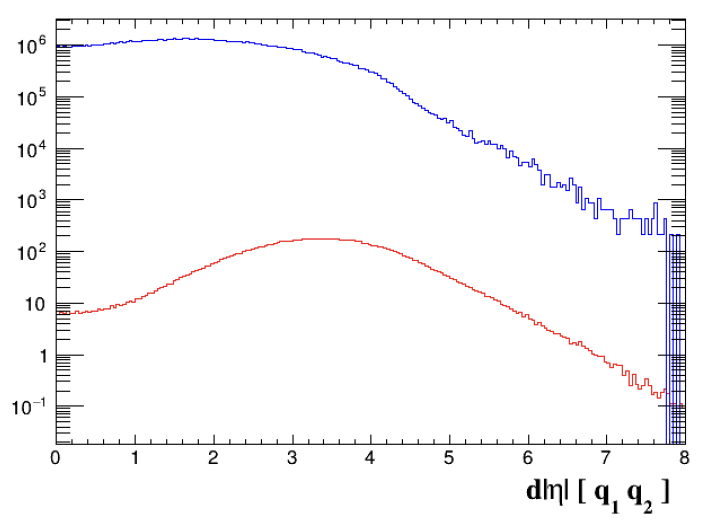}}
\end{center}
\caption{Anticipated number of events for the same sets of collisions as in Fig.~\ref{4}, again before cuts, as a function of $\left| \eta_d \right|$ for``vector difference'' of quark momenta (in notation of \cite{MadAnalysis}).}
\label{6}
\end{figure}

\begin{figure}
\begin{center}
\resizebox{\columnwidth}{!}
{\includegraphics{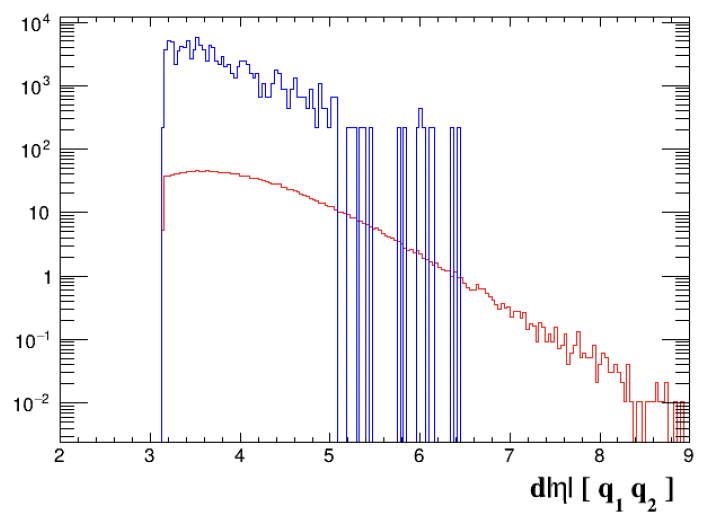}}
\end{center}
\caption{Anticipated number of events for the same sets of collisions as in Fig.~\ref{4}, but with the cuts of Fig.~{6}, as a function of $\left| \eta_d \right|$ for``vector difference'' of quark momenta (in notation of \cite{MadAnalysis}).}
\label{7}
\end{figure}

\begin{figure}
\begin{center}
\resizebox{\columnwidth}{!}
{\includegraphics{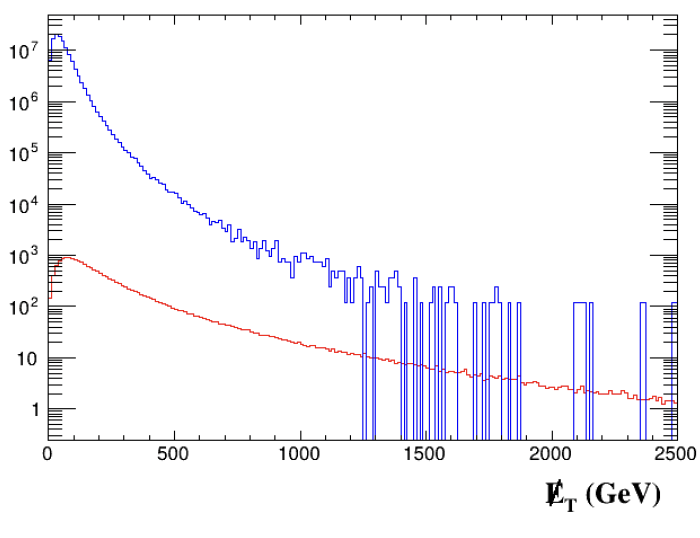}}
\end{center}
\caption{Before cuts, anticipated number of events for a pair of higgsons $h^0$ accompanied by a Z jet, as a function of ``missing transverse energy'' $\slashed{E}_T$. (Again, all observables are defined in \cite{MadGraph} and\cite{MadAnalysis}.) The neutrino background is orders of magnitude larger than the signal, but more concentrated toward lower energy, making this a natural choice for the first cut. (The relatively large size of the fluctuations in the background histograms is explained in the caption to Fig.~\ref{4}.) }
\label{8}
\end{figure}

\begin{figure}
\begin{center}
\resizebox{\columnwidth}{!}
{\includegraphics{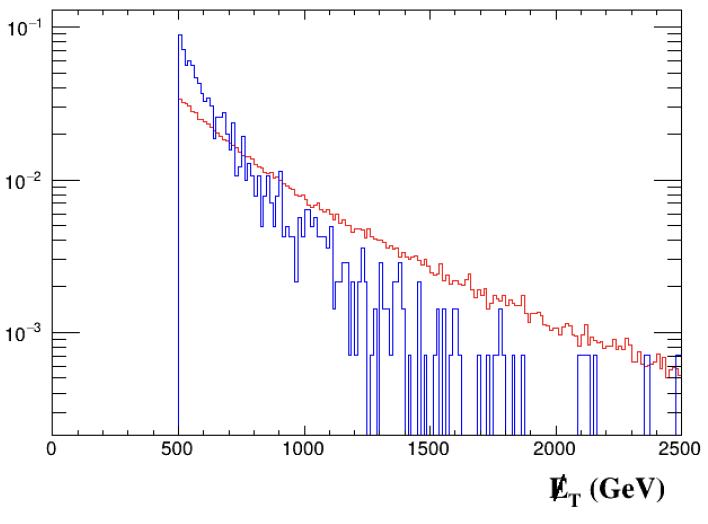}}
\end{center}
\caption{After the cuts specified in Table 2, anticipated number of events for higgson pair accompanied by Z jet, as a function of $\slashed{E}_T$ -- scaled to one in this plot to emphasize the difference in 
$\slashed{E}_T$ for signal and background. }
\label{9}
\end{figure}

We now give two examples of how the present particle can be discovered in proton colliders, with a small fraction 
of the results for observables shown in Figs.~\ref{4}-\ref{9}. 
\
\subsection{First example -- higgson pair accompanied by 2 quark jets, with 14 TeV center of mass energy and 500 fb$^{-1}$ integrated luminosity}

In Fig.~\ref{cross} it can be seen that the largest cross-section is for production of a higgson pair accompanied by a pair of quark jets. We therefore focus on this process in searching for optimized cuts at 14 TeV center of mass energy. The standard model background for this single process is a neutrino pair accompanied by two quark jets.

Below, $j_1$and $j_2$ are the two quark jets, $p_T$ is the transverse momentum, $\eta$ is the pseudorapidity, and other quantities are defined in \cite{MadGraph} and \cite{MadAnalysis}. The final optimized cuts (obtained after many one-million-event runs) are the following:

\noindent
Cut 1: $p_T$ of  $j_2$ $<  250$ GeV

\noindent
Cut 2: $ \left| \eta_{j_1} - \eta_{j_2} \right| > 3$

\noindent
Cut 3: $\left| \eta_s \right| > 1.5$ for ``vector sum'' of quark jet momenta 

\noindent
Cut 4: energy of $j_1$ $> 420$ GeV

\noindent
Cut 5: energy of $j_2$ $> 420$ GeV

\noindent
Cut 6: $\left| \eta_d \right| > 3.15$ for ``vector difference'' of jet momenta 

\noindent
Cut 7: sum of quark jet energies $> 1500$ GeV

\noindent
Cut 8: ``missing transverse energy'' $> 70$ GeV

\noindent
Cut 9: `invariant mass" $> 1500$ GeV

In Table 1 we show the number of signal and background events kept and rejected by each cut (with numbers rounded to the nearest integer).
\begin{tabular}{ |p{0.5cm}| |p{1.1cm}| p{1.3cm}| p{1.8cm}|  p{1.8cm}|  }
 \hline
  \multicolumn{5}{|l|}{Table 1: number of events kept or rejected, example 1} \\
 \hline
 cut  & signal kept & signal rejected & background kept & background rejected \\
 \hline
1  &  10307  & 13 & 101700400  & 116602100  \\
2 &   9521  & 785  & 27214465 & 74485935  \\
3 & 8848 &  673 & 19331560  & 7882905  \\
4  & 6224 & 2624 & 2939007 & 16392553  \\
5  &  4267  & 1956 & 869280 & 2069726 \\
6 & 3197 & 1070 & 397747 & 471533 \\
7  & 2236 & 961 & 160670 & 237076 \\
8 & 2002  & 234 & 121157 & 39512  \\
9  &  1946  & 56 & 113735  & 7422  \\
 \hline
\end{tabular}

\bigskip
Our calculated total cross-section for production of a $h^0$ pair accompanied by two quark jets is 20~fb at 14 TeV and 770~fb at 100 TeV.  With the optimized cuts of Table 1, the cross-section for the signal (higgson pair and quark jet pair) is reduced by a factor of about 5, but the background is reduced by a factor of about 2000. 

More specifically, for this integrated luminosity, after the cuts there are about 1950 signal events and 114,000 background events, so that the significance, according to the simple  prescription $n_s/\sqrt{n_b+n_s}$ used here and in \cite{MadGraph} and\cite{MadAnalysis}, is $> 5\sigma$. 

This is, in fact, our principal conclusion: If it exists, the present dark matter candidate can be detected at the high-luminosity LHC with optimized cuts. 

For higher integrated luminosities, extending up to 3000 fb$^{-1}$, the significance reaches $>5 \sigma$ with simpler cuts -- and
the cuts used here are, of course, far less sophisticated (and probably less elegant) than those which can be developed on a longer time scale, with a much more nuanced understanding of detector capabilities and limitations, as the time approaches for actual data-taking at the HL-LHC.

A calculation for the current LHC, with the same cuts but 13 TeV center-of mass-energy and and 100 fb$^{-1}$ integrated luminosity, resulted in a significance of less than 
$3 \sigma$, indicating that the HL-LHC is required for credible discovery of the present particle.

The cross-sections for higgson production in Fig.~\ref{cross} are an order of magnitude higher than those for VBF Higgs pair creation. We attribute this  to destructive interference between the several competing processes for the Higgs that can be seen in e.g. Fig.~1~(b) of ~\cite{VBF-2013}, plus the larger mass of the Higgs. 
As pointed out in~\cite{100TeV}, ``All the HH production mechanisms feature the interference between diagrams that depend on the self-coupling with diagrams that do not", and the cross-sections also decrease with increasing mass. 

However, the cross-section for $h^0$ production at 14 TeV is only about half the total cross-section for double Higgs production at this energy when all processes are included: approximately 40~fb~\cite{100TeV}.

The calculations reported here are leading-order, but the prediction at 14 TeV should be quantitative -- since the closely related LO and N$^3$LO cross-sections for VBF Higgs pair production at this energy were found to be nearly the same~\cite{VBF-N3LO}. At 100 TeV the results are presumably less reliable, and errors of a factor of 2 or 3 would not be surprising.

\subsection{Second example -- higgson pair accompanied by Z jet, with 100 TeV center of mass energy and 3000 fb$^{-1}$ integrated luminosity}
There are, of course, many other possible signals besides those in Fig.~\ref{cross} (and ordinarily many Feynman diagrams corresponding to a given signal). Figs.~\ref{8} and \ref{9} show our results (again obtained with MadGraph and MadAnalysis) for one low-order independent signal: a single Z jet accompanying a higgson pair. 

The cross-section is found to be a potentially achievable 6 fb before cuts. The standard model background is a Z jet accompanying a pair of neutrinos, with a cross-section of 40 pb before cuts.

The optimized cuts are:

\noindent
Cut 1: ``missing transverse energy'' $> 500$ GeV

\noindent
Cut 2: for Z pseudorapidity $-3.5 < \eta < 3.5$

In Table 2 we show the number of signal and background events kept and rejected by each cut (with numbers again rounded to the nearest integer).
\begin{tabular}{ |p{0.5cm}| |p{1.1cm}| p{1.3cm}| p{1.8cm}|  p{1.8cm}|  }
  \multicolumn{5}{}{Table 2: events kept or rejected, second example} \\
  \hline
  \multicolumn{5}{|l|}{Table 2: number of events kept or rejected, example 2} \\
   \hline
 cut  & signal kept & signal rejected & background kept & background rejected \\
 \hline
1  &  14965  & 3485 & 70835100  & 49389660  \\
2 &   2493  & 12472  & 169276 & 70665823  \\
 \hline
\end{tabular}

\bigskip
After the cuts of Table 2, there are about 2500 signal events and 169,000 background events, so that the significance, according to the simple  prescription stated above and used in \cite{MadGraph} and\cite{MadAnalysis}, is about $6\sigma$. This then provides an independent way to detect and identify the present particle if a 100 TeV proton collider can be achieved and can acquire 3000 fb$^{-1}$ of integrated luminosity.

\section{\label{sec:sec4}Production cross-sections and experimental signatures in electron-positron and muon colliders}

\begin{figure}
\begin{center}
{\includegraphics[width=0.55\linewidth]{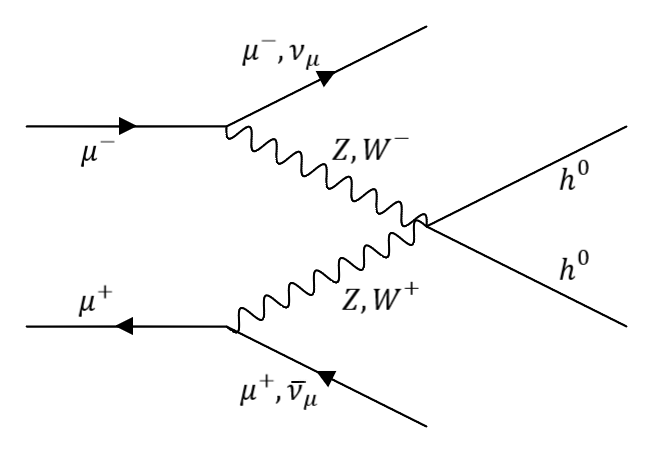}}
{\includegraphics[width=0.43\linewidth]{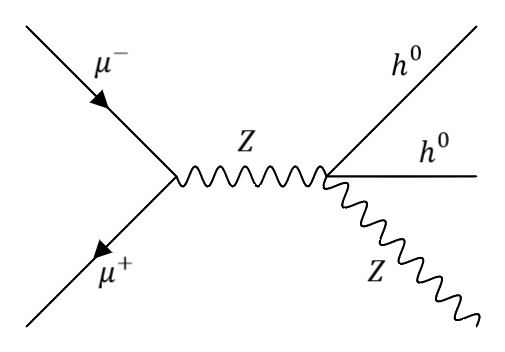}}
\end{center}
\caption{Production via vector boson fusion and Z-strahlung.}
\label{10}
\end{figure}

The other more powerful colliders planned for future decades include e$^+$e$^-$ colliders and a $\mu^+ \mu^-$  collider with center-of-mass energies up to 10 TeV. Here we will make no distinction between the lepton colliders because the results are nearly the same except at the lowest energies. Two simple processes are shown in Fig.~\ref{10}: vector boson fusion (VBF) and Z-strahlung. In our calculations, simple VBF with W bosons has a sizable cross-section at high energy -- about 700 fb at 10 TeV for polarized muon beams -- but this process is unobservable because only higgsons and neutrinos are produced. 
 
\begin{figure}[H]
\begin{center}
\resizebox{\columnwidth}{!}
{\includegraphics{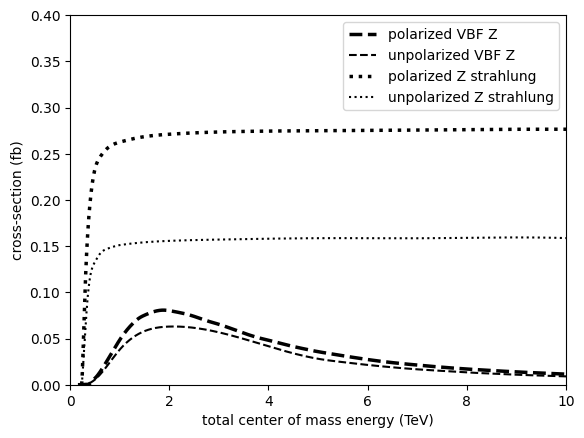}}
\end{center}
\caption{Cross-sections for the processes of Fig.~\ref{10}.}
\label{12}
\end{figure}

Our results for the simplest observable processes are shown in Fig.~\ref{12}. The cross-sections are unfortunately small -- below 0.3 fb even for polarized beams at the highest energies. This means that detection of the present particle at lepton colliders will be challenging.

\section{\label{sec:sec5}Cross-section for direct detection }

The lowest-order coherent processes for scattering of the present particle off a nucleus are represented by Fig. 4 of \cite{DM2021}. For a dark matter particle with initial velocity $v \ll c$ it is a good approximation to neglect both initial and final velocities. It is also a good approximation to neglect all but the valence $u$ and $d$ quarks. (In the familiar  scattering through Higgs exchange~\cite{Ellis,Mambrini}, all 6 quarks make comparable contributions, because the heavy quarks have stronger couplings. But here all quarks have comparable couplings, and the number density is relatively low even for strange quarks.) 

The interior loop for each amplitude, involving a given virtual quark with mass $m(p)$ (within a nucleon, at energy scale $p$) and a given vector boson (or its Goldstone boson) with mass $M$, involves
\begin{align}
\hspace{-0.5 cm}
\int \frac{d^4 p}{\left(2 \pi \right)^4} &\, \frac{-1}{\left( p^{\mu} p_{\mu} - M^2  +i \epsilon \right)^2} \, \frac{i \left( \slashed{p} + m(p) \right) } { p^{\mu} p_{\mu} - m(p)^2 +i \epsilon  } \nonumber\\
& \approx - i \int \frac{d^4 p}{\left(2 \pi \right)^4} \, \frac{1}{\left( p^{\mu} p_{\mu} - M^2  +i \epsilon \right)^2} \, \frac{m(p)} { p^{\mu} p_{\mu} +i \epsilon  } \label{eq1}
\end{align}
since the integral over $\slashed{p}$ is odd in $p_{\mu}$ and $m(p) \ll M$. Here $m(p)$ is the effective quark mass at momentum $p$. In the original right-hand factor the 4-momentum is $p'  = p + p_0 $, where $p_0$ is the momentum of the incident quark. But we can rename the momentum, with $p' \rightarrow p$ and $p \rightarrow p - p_0$ in the volume element and other factor. Then neglecting $p_0$ in this other factor introduces only a small error in the integral, and we obtain the form on the left side of (\ref{eq1}).

This loop integral is manifestly convergent and it
can be evaluated as a simple integral over the Euclidean momentum after a Wick rotation -- or alternatively using the residue theorem for $p_0$, followed by 3-dimensional integration. The result is
\begin{align}
I = \frac{1 }{16 \pi^2}\frac{m(M)}{M^2} \; \label{eq2}.
\end{align}

For each amplitude this integral is multiplied by a set of factors -- involving the metric tensor $g_{\mu \nu}$ (in the W or Z propagator), gamma matrices $\gamma^{\mu}$ and gauge generators (at vertices), and couplings proportional to the weak interaction coupling constant $g$ --  which turns out to reduce in each case to a simple overall factor 
$\lambda$  proportional to $g^4$ that is $\sim 0.02 - 0.1 $, depending on the diagram. (W$^+$W$^-$ and W$^-$W$^+$ diagrams must be combined.)

The effective interaction with a given initial quark $q$ in a nucleon $N$ is then $\lambda \, I \expval{\overline{q} q}{N}$, where $\expval{\overline{q} q}{N}$ is the expectation value of the number of these quarks in the nucleon. Notice that, in $I$, $m(p)$  is the mass of the central virtual quark -- e.g., a $u$ if the initial and final quarks are $d$ and the vector bosons are $W^{\pm}$.

For Higgs exchange in the case of a hypothetical scalar particle having a Higgs-boson coupling $\lambda_{\chi} v$, where $v$ is the Higgs-field vacuum expectation value, the effective interaction with a given quark, having effective mass $m_q$ and effective Yukawa coupling $y_q = m_q/v$, 
is $\expval{\overline{q} \, q \, y_q}{N} _{eff}  \lambda_{\chi} v / M_H^2 =  \lambda_{\chi} \expval{\overline{q} q \,m_q}{N} _{eff} /M_H^2 $, where $M_H$ is the Higgs mass.

Then the difference between the present loop process and Higgs exchange is that the amplitude is reduced by a factor of
\begin{align}
&\frac{6 \left( 16 \pi^2 \right)^{-1} \left[ \lambda \expval{\overline{q} q}{N} m (M) / M^2 \right]_{val,avg}}
{6 \, \lambda_{\chi}\left[ \expval{\overline{q} q \, m_q }{N}   \right]_{eff,avg} / M_H^2 } \nonumber \\
& \hspace{1cm} = \frac{1}{16 \pi^2 }  \frac{M_H^2}{M_{avg}^2}  \frac{ \left[ \lambda  \expval{\overline{q} q}{N} \,m (M) \right]_{val,avg}}
{\lambda_{\chi} \left[  \expval{\overline{q} q \, m_q }{N}   \right]_{eff,avg}}
\end{align}
where $\left[ \cdots \right]_{val,avg}$ is an average over the valence quarks within a nucleon $N$ and their interactions with W and Z bosons, 
$\left[ \cdots \right]_{eff,avg}$ is the average of the effective value of the product over all 6 quarks, the averages are also over both nucleon species, and $M_{avg}$ is an average over W and Z masses. 

Since $(M_H^2/M_{avg}^2) \, \lambda \sim 0.1$, if we had $ \left[ \expval{\overline{q} q}{N} m (M) \right]_{val,avg} \approx \left[  \expval{\overline{q} q \, m_q}{N}   \right]_{eff,avg}$, the amplitude for the present process would be lowered only by about $0.1 \left(16 \pi^2 \right)^{-1}/\lambda_{\chi}  \sim 10^{-3}/\lambda_{\chi} $. 
But the appropriate value of $\expval{\overline{q} q}{N}$ for a valence quark is roughly $1/6$, and the appropriate values of $m(p)$ at high energy are the current (``free quark'') masses, which are about 2.4 MeV and 4.8 MeV for up and down quarks respectively, so that $ \left[ \expval{\overline{q} q}{N} m (M) \right]_{val,avg} \sim 0.0005 $ GeV, whereas $\left[  \expval{\overline{q} q \, m_q}{N} \right]_{eff,avg} \sim 0.05$ GeV~\cite{Ellis,Mambrini}. The amplitude for the present loop process is then reduced, relative to that for Higgs exchange, by a further factor of $\sim 10^{-2} $, or by $\sim 10^{-5}/\lambda_{\chi} $ overall, implying that the cross-section is lowered by $\sim 10^{-10}/\lambda_{\chi}^2 $. Since $\lambda_{\chi} = 0.01/4$ for a particle of mass 100 GeV would imply a Higgs-exchange cross-section of $\sim 10^{-46}$ cm$^2$~\cite{Mambrini}, for the present process this will be lowered to $\sim  10^{-51}$ cm$^2$.

The cross-section for direct detection of the present dark matter candidate is then $\sim 10^{-51}$ cm$^2$, which is below the sensitivities of current and planned detectors and  well within the neutrino fog. This estimate is confirmed by detailed calculation within a parton description, but an accurate result would require sophisticated treatment of form factors, parton distribution functions, higher-order processes, and other issues involving interactions for a loop process inside a nucleus.

If $m(p)$ in (\ref{eq1}) and (\ref{eq2}) were taken to be the mass of a constituent quark, the predicted cross-section would be above $10^{-48}$ cm$^2$ and within range of current experiments~\cite{XENON,LZ,PandaX}. But the appropriate mass at high energy (and a short time scale) is instead the current mass, which is much smaller, reducing the cross-section by orders of magnitude.

\section{\label{sec:sec6}Conclusions}

Based on the calculations above, we conclude that the present dark matter WIMP is consistent with all current experimental and observational constraints, and 
(1)~its mass is about 70 GeV if it is the dominant dark matter particle (although somewhat larger in a multicomponent scenario); (2)~the annihilation cross-section and mass are consistent with those inferred from gamma-ray and antiproton observations based on a dark matter interpretation; (3)~it should be detectable in the high-luminosity LHC with $> 5 \, \sigma$ statistical significance; (4)~it should be observable with additional signatures in more powerful future proton colliders; (5)~separating signal from background will be challenging in an e$^+$e$^-$ or muon collider; (6)~direct detection will also be challenging, since the cross-section is $\sim 10^{-51}$ cm$^2$.

Our principal conclusion is that this particle should be detectable at the high-luminosity LHC, perhaps after only two years with an integrated luminosity of 500 fb$^{-1}$. 

It can also be observed and identified in further astrophysical observations, in the experiments planned for future colliders, and possibly through direct detection with extended capabilities such as directional detection.

\end{document}